\documentclass[preprint2]{aastex}

\shorttitle{Peaks in the CMB}
\shortauthors{Netterfield et al}

\begin{document}

\title{A measurement by BOOMERANG of multiple peaks in the angular power
spectrum of the cosmic microwave background}

\author{
C.B.~Netterfield\altaffilmark{1}, 
P.A.R.~Ade\altaffilmark{2},
J.J.~Bock\altaffilmark{3},
J.R.~Bond\altaffilmark{4},
J.~Borrill\altaffilmark{5},
A.~Boscaleri\altaffilmark{6},
K.~Coble\altaffilmark{7},
C.R.~Contaldi\altaffilmark{4},
B.P.~Crill\altaffilmark{8},
P.~de Bernardis\altaffilmark{9},
P.~Farese\altaffilmark{7},
K.~Ganga\altaffilmark{10},
M.~Giacometti\altaffilmark{9},
E.~Hivon\altaffilmark{10},
V.V.~Hristov\altaffilmark{8},
A.~Iacoangeli\altaffilmark{9},
A.H.~Jaffe\altaffilmark{11},
W.C.~Jones\altaffilmark{8},
A.E.~Lange\altaffilmark{8},
L.~Martinis\altaffilmark{9},
S.~Masi\altaffilmark{9},
P.~Mason\altaffilmark{8},
P.D.~Mauskopf\altaffilmark{12},
A.~Melchiorri\altaffilmark{13},
T.~Montroy\altaffilmark{7},
E.~Pascale\altaffilmark{6},
F.~Piacentini\altaffilmark{9},
D.~Pogosyan\altaffilmark{4},
F.~Pongetti\altaffilmark{9},
S.~Prunet\altaffilmark{4},
G.~Romeo\altaffilmark{14},
J.E.~Ruhl\altaffilmark{7},
F.~Scaramuzzi\altaffilmark{9}
}

\affil{
$^{1}$ Depts. of Physics and Astronomy, University of Toronto, Canada \\
$^{2}$ Queen Mary and Westfield College, London, UK \\
$^{3}$ Jet Propulsion Laboratory, Pasadena, CA, USA \\
$^{4}$ Canadian Institute for Theoretical Astrophysics, 
		University of Toronto, Canada \\
$^{5}$ National Energy Research Scientific Computing Center, 
		LBNL, Berkeley, CA, USA \\
$^{6}$ IROE-CNR, Firenze, Italy \\
$^{7}$ Dept. of Physics, Univ. of California, 
		Santa Barbara, CA, USA \\
$^{8}$ California Institute of Technology, Pasadena, CA, USA \\
$^{9}$ Dipartimento di Fisica, Universita' La Sapienza, 
		Roma, Italy \\
$^{10}$ IPAC, California Institute of Technology, Pasadena, CA, USA \\
$^{11}$ Department of Astronomy, Space Sciences Lab and
		Center for Particle Astrophysics, \\ 
		University of CA, Berkeley, CA 94720 USA\\
$^{12}$ Dept. of Physics and Astronomy, Cardiff University, 
		Cardiff CF24 3YB, Wales, UK \\
$^{13}$ Nuclear and Astrophysics Laboratory, University of Oxford, 
		Keble Road, Oxford, OX 3RH, UK\\
$^{14}$ Istituto Nazionale di Geofisica, Roma,~Italy \\
}

\begin{abstract}
This paper presents a measurement of the angular power spectrum of the
Cosmic Microwave Background from $\ell=75$ to $\ell=1025$ ($\sim 10'$ to
$2.4^o$) from a combined analysis of four 150 GHz channels in the
BOOMERANG experiment.  The spectrum contains multiple peaks and minima,
as predicted by standard adiabatic-inflationary models in which the
primordial plasma undergoes acoustic oscillations.  These results, in
concert with other types of cosmological measurements and theoretical
models, significantly constrain the values of $\Omega_{\rm tot}$,
$\Omega_{b}h^2$, $\Omega_{c}h^2$ and $n_s$.
\end{abstract}
\keywords{Cosmic Microwave Background Anisotropy, Cosmology}

\section{Introduction}

The presence of a harmonic series of ``acoustic'' peaks in the angular
power spectrum of the cosmic microwave background (CMB) was predicted
as early as 1970 \citep{SZ70,Peb70}.  These
peaks arise due to the evolution of pressure waves in the pre-recombination
universe  and  are a generic feature \citep{Bon87} of most, but not all, cosmological
models (e.g., \citet{Hu1997}).  Specifically, a well-defined set of peaks
is predicted for both adiabatic and some classes of isocurvature models
of structure formation, but not for models that rely on topological
defects.  

Since the COBE measurement of the amplitude of fluctuations in the
cosmic microwave background at the largest scales \citep{bennet}, a
large literature has developed which predicts, in the context of
adiabatic cold dark matter (CDM) models, the relative position and
amplitude of these peaks for different values of the fundamental
cosmological parameters.  A general prediction of these models is the
presence of a dominant fundamental peak at an angular scale $\approx
1^o$ ($\ell\approx 200$), decreasing in angular scale when $\Omega$
decreases.  Data from a variety of experiments
\citep{Mil99, Maus2000, Hana2000} including a small fraction of the data
from the BOOMERANG 1998/1999 Long Duration
Ballooning (BOOM/LDB) campaign (\citet{debe2000}; B00 hereafter) clearly
show this feature and provide strong evidence for a low curvature
universe, a generic prediction of many inflation models.

There is also convincing evidence that the broad-band average of power
at smaller angular scales gradually declines in a manner consistent with
adiabatic CDM models \citep{Pad01, church}.  However, these experiments
do not have the necessary combination of sensitivity and sky coverage to
convincingly detect or reject the presence of harmonics of the
fundamental peak in the power spectrum.  The detection of such harmonic
peaks would provide strong evidence
that the scenario of structure formation from acoustic oscillations in
the primordial plasma is accurate.

Present here is an analysis of a larger set of data than previously
released from the BOOM/LDB \citep{debe2000} experiment which shows clear
evidence of multiple peaks in the angular power spectrum of the CMB.
Data from four separate detectors that each observe 1.8\% of the sky are
combined.  A new data analysis algorithm is used and refined estimates
of the beam shape and overall experimental calibration are presented.
The spectrum is consistent with low spatial curvature,
$\Lambda$-dominated adiabatic CDM models.

\section{Instrument}

BOOMERANG is a Long Duration Balloon (LDB) experiment designed to
measure the angular power spectrum of the CMB at degree and sub-degree
scales.  For a complete description of the instrument see
\citet{Cril2001} and \citet{Piac2001}.
Instrument characteristics are summarized in
Table~\ref{table_inst}.


\begin{deluxetable}{cccc}
  \tabletypesize{\scriptsize} 

  \tablecaption{Instrument Characteristics. \label{table_inst}} 
  \tablewidth{0pt} 
  
  \tablehead{ 
\colhead{Channel} & \colhead{Band (GHz)} &
  \colhead{$NET_{CMB}$ ($\mu K \sqrt{s}$)} & \colhead{Beam FWHM (')} }

  \startdata
   B150A  & 148.0 - 171.4 & $130$  & $9.2\pm 0.5$\\
   B150B  & 145.8 - 168.6 & Variable & $9.2\pm 0.5$\\
   B150A1 & 145.5 - 167.3 & $231$  & $9.7\pm 0.5$\\
   B150A2 & 144.0 - 167.2 & $158$  & $9.4\pm 0.5$\\
   B150B1 & 144.2 - 165.9 & $196$  & $9.9\pm 0.5$\\
   B150B2 & 143.7 - 164.3 & $184$  & $9.5\pm 0.5$\\
  \tableline
   90 (2 Chs)   &  79  -  95  & $140$  & $18\pm 1  $\\
   240 (3 chs) & 228  - 266 & $200$  & $14.1\pm 1$\\
   410 (4 chs) & 400  - 419 & $\sim 2700$ & $12.1\pm 1$\\
  \enddata
  
  \tablecomments{Summary of relevant instrument characteristics.  Only
  results from the 150GHz channels are presented in this paper.  B150B
  is not used due to non-stationary detector noise. The bandwidth limits
  are computed to include 68\% of the total detected power for a flat
  spectrum source. The NET is computed at 1Hz.}
\end{deluxetable}

BOOMERANG was launched for its first LDB flight on December 29, 1998
from McMurdo station, Antarctica, and acquired 257 hours of data from an
altitude of $\approx39$km.  BOOMERANG is comprised of a 1.2m off axis
parabolic mirror which feeds a cryogenic mm-wave bolometric receiver.
Observations are made simultaneously in four unpolarized bands centered
at 90 GHz, 150 GHz, 240 GHz and 410GHz.

The telescope is steerable in azimuth by moving the entire gondola, and
in elevation by moving an inner frame containing both the receiver and
the optics.  The illumination of the optics is not modulated by the
scan, which minimizes scan synchronous optically generated offsets.
Extensive shielding permits observations in the azimuth range $\pm 60^o$
from the anti-sun direction, for all sun elevations experienced in the
antarctic ballooning environment.

\section{Observations and Calibration}

\begin{figure}[tbhp]
  \epsscale{0.7} \plotone{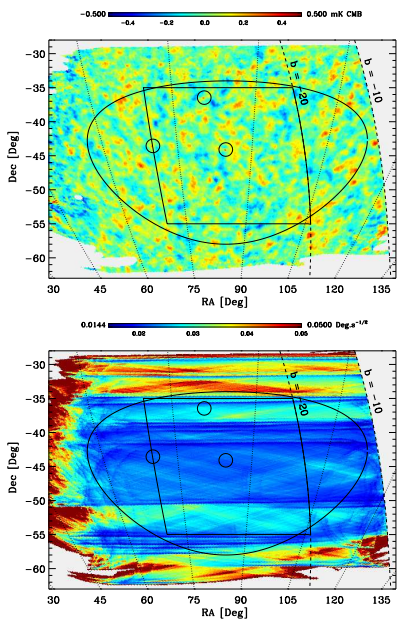} 
  \caption{Sky coverage.  The upper panel shows the BOOMERANG 150GHz
  map.  The locations of the three bright quasars are circled.  The sky
  subset used in B00 (rectangle) and this paper (ellipse) are shown.
  The bottom panel shows the integration time/pixel.  The ellipse for
  this analysis was chosen to include the well sampled sky, and to avoid
  the poorly sampled sky} \label{fig:coverage}
\end{figure}

Observations are made by scanning the telescope in azimuth by $60^o$
peak-to-peak at an angular velocity of $2 ^o/{\rm s}$ (hereafter 2dps)
(for the first half of the flight) or $1^o/{\rm s}$ (hereafter 1dps)
(for the second half of the flight) at fixed elevation.  Each day, the
elevation is shifted.  Observations are made at elevations of $40 ^o$,
$45 ^o$, and $50 ^o$.  The scans are centered well away from the
Galactic plane.  Interspersed in the CMB observations are observations
of selected point sources near the Galactic plane.  The CMB sky coverage
is shown in Figure~\ref{fig:coverage}.


The results in this paper contain data from the entire BOOM/LDB flight (as
compared to the second half only in B00)
and incorporate data from 4 150 GHz channels (as opposed to 1
for B00).  This was enabled by using the new data analysis techniques
described in Section~\ref{sec:DataAnalysis} and in \citet{Hivo2001}.
In addition, the most significant source of systematic uncertainty at
multipoles, $\ell>400$ is the effective beam size (dominated by pointing
uncertainties).  Since the release of B00, the understanding of the
beam and pointing has been significantly improved, allowing computation
of the power spectrum out to higher multipoles.

Observations of extra Galactic sources in the main map and scanned
observations of bright HII regions near the Galactic plane are used to
estimate the beam full width half maxima (FWHM) for each channel. These
values are presented in Table \ref{table_inst}.  While these
observations provide good statistics on the width of the main lobe, the
signal to noise of these data are not sufficient to fully characterize
the near sidelobe response and deviation from gaussianity of the beam.
To obtain a detailed model of the near sidelobe response of the
telescope, a physical optics calculation of the beam shape is performed
for each channel using the measured position of the horns in the focal
plane.  To check the precision of the model calculations, a comparison
with near field beam maps is made.  Azimuthal cuts through the telescope
beams are measured to the $\sim -20$ dB level.  While the channels
utilizing conical feed horns compare well with the beam map data, a
$\sim 10\% $ discrepancy is observed between the calculated and measured
FWHM of the photometer beams.  In all cases, the measured FWHM are
larger than the calculated beam size.  This discrepancy is attributed to
the multi--moded nature of the Winston concentrators utilized by these
channels.  For these channels the FWHM are scaled to fit the
measurements of RCW38. To correct for the extended nature of the source,
the angular extent of RCW38 was measured independently by the ACBAR
instrument to be $2.5'$ \citep{acbar}.  The two dimensional far-field
radiation patterns from the physical optics calculation are then used to
generate the window functions for each channel.

The telescope pointing is determined using a combination of
rate-gyros and an azimuth sun sensor.  In order to recover the long time
scale pendulations, the gyroscopes are integrated with a 400s time
constant.  Given the quoted noise in the gyroscopes of less than
$8'/\sqrt{hour}$, a pointing uncertainty of less than 2.7' ($1 \sigma$)
is predicted.  The Galactic plane point source observations give a
pointing uncertainty of 2.5' ($1 \sigma$).  In the analysis, the
calculated beam is convolved with the Gaussian approximation of the
pointing uncertainty.  

The pointing solution has been improved in this analysis (compared with
B00) by better use of the pitch and roll rate gyros, and the regression
of a thermally dependent offset in the elevation.  A re-analysis of the
pointing jitter in the B00 pointing solution, utilizing the apparent
centroids of point sources along the Galactic plane yields an effective
beam size of $12.9 \pm 1.4'$, for the original pointing solution (as
compared to the quoted $10'\pm 1'$ used in B00).  When the calibration
and beam uncertainties are taken into account, the new results lie, bin
by bin, within the overall uncertainty of the B00 spectrum, which was
restricted to $\ell$ </= 600.  The B00 results are systematically lower
at high $\ell$ than those presented here, due to the smaller effective
beam that was assumed for the B00 analysis.  However, it is reassuring
to note that correcting the B00 spectrum with the new estimate of the
B00 pointing jitter eliminates any residual discrepancy between the new
results and the B00 spectrum \citep{deb01}.

The gain calibrations of the 150 GHz channels are obtained from
observations of the CMB dipole. The data are high-pass filtered with a
filter described by $F=0.5\left(1-\cos\left(\pi\nu\over
f_0\right)\right)$ for $0<\nu<f_0$ and 1 elsewhere. In order to retain
more large-scale information than is needed in the anisotropy analysis,
$f_0$ is set to the relatively low value of 0.01~Hz.  To compare with
the data, we artificially sample the CMB dipole signal \citep{Line1996},
corrected for the Earth's velocity around the sun \citep{Stump1980},
according to the BOOMERANG scanning, and filter this fake time stream in
the same way as the data.  The 1dps data is then fit simultaneously to
this filtered dipole, a similarly filtered dust emission model
\citep{Fink99}, an offset and the BOOMERANG 410 GHz data for all data
more than $20^o$  below the Galactic plane. The dipole calibration
numbers obtained with this fit are robust to changes in Galactic cut,
and to whether or not a dust model is included in the fit; this
indicates that dust is not a serious problem for the contamination. They
are insensitive to the inclusion of a 410 GHz channel in the fit,
which is a general indication that there is no problem with a wide range
of systematics such as atmospheric contamination, as these would be
traced by the 410 GHz data.

Overall, the calibration of the spectrum has risen by 10\% in $C_l$ (5\%
in $\Delta T$) compared to B00 due to a refinement of the dipole calibration
(improved treatment of the time stream filters) and is further raised by a
better calculation of the beam sidelobes.

\section{Data Analysis} \label{sec:DataAnalysis}

The data are analyzed in four basic steps: i) the reduction of the raw
data into pointed and flagged time streams, ii) the estimation of the
noise via an iterative map-making algorithm, iii) the estimation of the
angular power spectrum via Monte-Carlo calibrated spherical harmonic
transforms, and iv) estimation of parameters by likelihood methods.  The
size and nature of the BOOMERANG data  have required the development
of new techniques.

In the reduction of the raw bolometer data, the filter response of the
detector and associated electronics is deconvolved from the time stream,
and transient phenomena (predominantly cosmic rays) are flagged and replaced in
the time stream with a constrained realization of the noise.  Similarly,
the RA/Dec pointing for each channel is reconstructed from the rate
gyros, azimuth sun sensor angle, GPS co-ordinates of the package, and
the focal plane geometry.  Details can be found in \citet{Cril2001}.

The receiver noise for each channel is estimated from the raw time
stream by iteratively solving simultaneously for the detector noise
spectrum $n(f)$ and the maximum likelihood CMB map, $\Delta = (\bf {P}^\dagger
\bf {N}^{-1}\bf {P})^{-1}\bf {P}^\dagger \bf {N}^{-1} d$.  The algorithm
used is an approximate Jacobi method:

\vskip 0.5cm
\parbox[c]{8cm}{
{\bf loop on j}
\begin{itemize}
\item{$ \bf {n}^{(j)} = \bf{d} - \bf{P\tilde\Delta}^{(j)} \Rightarrow
\bf{N}^{(j)-1} =
\langle\bf{nn}^\dagger\rangle^{-1} $}
\item{$ \bf{\tilde\Delta}^{(j+1)} - \bf{\tilde\Delta}^{(j)} =
\left(\bf{P^\dagger P}\right)^{-1}\bf{P^\dagger}\bf{N}^{(j)-1}\bf{n}^{(j)}$}
\end{itemize}
{\bf end loop}}
\vskip 0.5cm

where is $\bf {N}$ is the time-time correlation matrix, $\bf {P}$ 
is the pointing
matrix, $\bf {d}$ is the time-stream data, and $\bf {n}$ is the noise time
stream.  If the noise is stationary, then $\bf {N}^{-1}$ is diagonal in 
Fourier
space, and multiplication by $\bf {N}^{-1}$ is just a convolution.  And
multiplication by $\left(\bf{P^\dagger P}\right)^{-1}\bf{P^\dagger}$ represents
binning into pixels and dividing by hits per pixel.  For details see
\citet{Prun2000}.  A complete map and noise spectra takes about 15
minutes on an alpha-ev67 computer.

Two closely related estimators were used to recover the underlying CMB
power spectrum $C_{\ell}$ from the data. Both methods are based on the 
Monte Carlo Spherical Harmonic Transform ({\sl MASTER}) technique described in
\cite{Hivo2001}. {\sl MASTER} allows fast and accurate determination
of $C_\ell$ {\em without} performing the extremely time
consuming matrix-matrix manipulations that characterize exact methods
and limit their applicability (\cite{Borr1999}). It can be
summarized as follows.  The spherical harmonic transform of a naively
binned map of the sky is calculated using a fast ${\cal
O}(N_{pix}^{1/2}\ell)$
method based on the Healpix tessellation of the sphere
\citep{Gors1998}. The angular power in the noisy maps, $\tilde{C_{\ell}}$,
can
be related to the true angular power spectrum on the full sky,
$C_{\ell}$, by the effect of finite sky
coverage ($M_{\ell \ell'}$), time and spatial filtering of the maps
($F_{\ell}$),
the finite beam size of the instrument ($B_{\ell}$), and instrument noise
(${N_{\ell}}$) as

\begin{equation}
\left\langle \widetilde{C_{\ell}}\right\rangle =
  \sum _{\ell'}M_{\ell\ell'}F_{\ell'}B_{\ell'}^{2}\left\langle
C_{\ell'}\right\rangle
+ \sum _{\ell'}M_{\ell\ell'}F_{\ell'}\left\langle N_{\ell'}\right\rangle.
\label{eqn_cl}
\end{equation}
 
The coupling matrix $M_{\ell \ell'}$ is computed analytically,
$B_{\ell}$ is determined by the measured beam,  $F_{\ell}$
is determined from Monte-Carlo simulations of signal-only time streams,
and $N_{\ell}$ from noise-only simulations of the time streams. 

The simulated time streams are created using the actual flight pointing
and transient flagging. The signal component of the simulated time
streams is generated from simulated CMB maps and the noise component
from realizations of the measured detector noise $n(f)$. $F_{\ell}$
and $N_{\ell}$ are determined by averaging over $150$ and
$200$ realizations respectively. Once all of the components are known the
estimation is carried out in two ways. 

In the first case the power is determined by solving directly for the
unbiased estimator $C_{\ell}$ of eqn.~\ref{eqn_cl}.  The uncertainties
in the estimated top-hat binned $C_{\ell}$ spectrum are measured by
averaging over ensembles (typically $400$ realizations) of signal+noise
simulations created using a best fit model power spectrum obtained from
the data.  This allows one to calculate the quantities needed to
approximate the full likelihood function for the binned $C_{\ell}$,
using the formalism of
\cite{bjk00}.

In the second case a suitable quadratic estimator of the {\em full sky}
spectrum in the {\em cut sky} variables ${\tilde C}_{\ell}$ together
with it's Fisher matrix is constructed via the coupling matrix $M_{\ell
\ell'}$ and the transfer function $F_{\ell}$ \citep{bjk98,b01}. The
underlying power is recovered through the iterative convergence of the
quadratic estimator onto the maximum likelihood value as in standard
Maximum Likelihood (ML) techniques. A great simplification and speed-up
is obtained due to the diagonality of all the quantities involved,
effectively avoiding the ${\cal O}(N^3)$ large matrix inversion problem
of the general ML method.  The extension of the quadratic estimator
formalism to montecarlo techniques such as MASTER have the added
advantage that the Fisher matrix characterizing the uncertainty in the
estimator is recovered directly in the iterative solution and does not
rely on any potentially biased signal+noise simulation ensembles.

The two procedures agree to within a few percent in the estimated values
with the quadratic estimator giving slightly more optimal errors (at the
$5\%$ level) over the sample variance limited range of the data. The
parameter extraction pipeline was run over results from both methods and
the two were found to agree to within the numerical accuracy of the
fits.

The drawback of using naively binned maps in the pipeline is that the
aggressive time filtering completely suppresses the power in the maps
below a critical scale $\ell_c$ \cite{Hivo2001}. This results in one or
more bands in the power spectrum running over modes with no power and
which are thus unconstrainable. In pratice we deal with this by binning
the power so that all the degenerate modes lie within the first band
$2<\ell<50$ and regularize the power in the band at DMR power in the
likelihood analysis. The estimate in the second band $50\ge\ell<100$
will be correlated to this regularized value and as such may be also be
considered to be biased by a prior theoretical input. We therefore
discard the estimates in the first two bands thus avoiding any
correlation to the regularizing scheme used to constrain the power on
the largest scales.

An area equivalent to $1.8\%$ of the sky was analysed. The region is
enclosed in an ellipse with $20$ and $12$ degree semi-axes centred at
$RA=85$ and $Dec=-46$. The data and simulations were pixelised with $7$
arcminute pixels (Healpix $N_{\rm side}=512$). The simulations were run at
an angular resolution up to $\ell=1300$.

Inspection of the BOOMERANG maps shows faint stripes of nearly constant
declination, (hereafter ``isodec'' strips) which vary in amplitude and
phase between bolometer channels.  The striping patterns vary on day
time scales, and are not reproduced in simulated maps made with the same
scan pattern and best estimated noise correlations from the time stream
data.

To eliminate this contaminant, all modes with a small $k_{RA}$ (which
corresponds to isodec stripes) are removed with a Fourier filter.  While
this clearly eliminates isodec stripes, it also filters out CMB signal.
This is accounted for by applying the same filter to the simulated maps
in the {\sl MASTER} procedure, so that the effects are included in the
determination of $F_{\ell}$.  The removal of the stripes still permits
an unbiased estimate of the power spectrum of the sky, but does cause a
considerable increase in the uncertainties at large angular scales.

The inclusion of several channels is achieved by averaging the maps
(both from the data, and from the Monte-Carlos of each channel)
before power spectrum estimation.  Weighting in the addition is by
hits per pixel, and by 
receiver noise at 1Hz.  Each channel has a slightly different beam size,
which must be taken into account in the generation of the simulated
maps.  Any inaccuracy in assuming a common $B_{\ell}$ in the angular
power spectrum estimation is then absorbed into $F_{\ell}$. 

The calculation of the full angular power spectrum and covariance matrix
for the four good 150 GHz channels of BOOMERANG (57103 pixels and
$\approx 216,000,000$ time samples) takes approximately 1 day running on
8 AMD-athlon 800 MHz work stations.

\section{Internal Consistency Tests}

The BOOMERANG observing strategy allows for a rich set of internal
consistency checks, implemented as a variety of difference maps in which
the sky signal should cancel.  The power spectra of these difference
maps are sensitive to improper characterization of the receiver noise,
and contamination not fixed to the celestial sphere.  The precision of
these difference tests are much more powerful than a comparison of the
power spectra, since the sample variance contribution to the power
spectrum error bars is proportional to the signal found in each bin,
which is near zero for the difference maps.  The results of these tests
are summarized in Table~\ref{table:chi2}.

\begin{deluxetable}{ccc}
\tabletypesize{\scriptsize}
\tablecaption{Internal Consistency Tests. \label{table:chi2}}
\tablewidth{0pt}
\tablehead{
\colhead{Test} & \colhead{Reduced $\chi^2$}   & \colhead{$P_>$}
}
\startdata
B150A 1dps - 2dps	& 0.91	& 0.57 \\
B150A1 1dps - 2dps	& 0.92 	& 0.56 \\
B150A2 1dps - 2dps	& 1.04	& 0.41 \\
B150B1 1dps - 2dps	& 2.73	& $7\times 10^{-5}$\\
B150B2 1dps - 2dps	& 0.60  & 0.91 \\
4 Ch 1dps - 2dps	& 1.80 	& 0.02 \\
4 Ch Left - Right	& 1.21	& 0.24 \\
(A+A1) - (A2+B2)	& 0.61	& 0.90 \\
\enddata

\tablecomments{Reduced $\chi^2$ with 19 degrees of freedom for internal
symmetry tests for BOOMERANG. $P_>$ gives the probability of obtaining a
$\chi^2$ larger than the one reported. B150B1 fails the test, and is not
used in the analysis.  The '4 Ch' entries combine maps from B150A,
B150A1, B150A2, and B150B2.  The 1dps-2dps 4 Ch spectrum fails
marginally.  This is dominated by 4 bins centered between $l=150$ and
$l=300$.  The mean signal of these 4 bins is $50\mu K^2$, compared to a
signal over the same range of 
$\approx 4000\mu K^2$(see Table~\ref{table:spec}).}

\end{deluxetable}

The most powerful of these tests is to difference the map made from data
acquired while scanning at 2dps (the first half of the flight) from data
acquired while scanning at 1dps. (the second half of the flight).  This
test is sensitive to solar and ground pickup, since between the center
of the 2dps data and the center of the 1dps data, the sun
moves $5^o$ on the sky, and the gondola has moved half way around the
continent between the time centroids of the two maps. The test is also
sensitive to  errors in the deconvolution of the transfer
function of the time-domain signal, since the scan speed changes the
spatio-temporal mapping of the signals.

This test is performed on each of the five 150 GHz channels
individually.  Without filtering out the isodec modes as described
previously this test is failed.  With the filtering, 4 of the 5 channels
pass, and 1 of the 5 channels (150B1) shows a small but statistically
significant signal.  This channel is excluded from subsequent
analysis. The isodec removal is applied to all of the channels included
in the analysis.

The 1st half - 2nd half difference test is also performed on maps with
the four channels combined.  At $l < 300$, there is a statistically
significant residual in the difference map at the level of $50\mu K^2$.
Since the signal only appears in the combined channel 1dps - 2dps
analysis, this is consistent with a noise term which changes slowly on
the sky and is correlated between channels, such as atmosphere.  At
these angular scales, the CMB signal is $\approx 4000\mu K^2$.  Since
the residual signal is small compared to the CMB signal, its effects are
neglected.

A test for artifacts specific to particular detectors is made by
differencing the map made from combining B150A and B150A1 with the
combination of B150A2 and B150B2, and a test for artifacts due to
scan-synchronous baselines is made by differencing maps using only the
left-going and right-going portions of the scans.  There is no evidence
of any residual signal in either of these tests, which is again
consistent with a noise term that changes slowly on time scales
comparable to the scan time.

\section{Foregrounds}

The comparison of the maps at the 4 different frequencies measured by
BOOMERANG is a powerful tool to test for contamination from foregrounds
at 150 GHz. At the resolution frequencies probed by BOOMERANG, thermal
emission from interstellar dust grains is expected to be the most
important foreground (see e.g. \citet{Tegm00}).  \citet{Mas01}
probes the level of dust in the BOOMERANG maps by correlating
BOOMERANG data with a dust template derived from the 3000 GHz
IRAS/DIRBE maps \citep{Schl99, Fink99} and extrapolating the
dust dominated 410 GHz signal to 150 GHz using the measured correlations.
The deduced power spectrum of dust fluctuations contributes less than
$1\%$ to the power spectrum of sky temperature measured at 150 GHz.
For this reason Galactic dust contamination is neglected in the following.

Radio point sources are another potential form of contamination in the
maps and angular power spectrum.  The \citet{wombat} radio point source
extrapolations are used to estimate the effects of known radio sources
in the BOOMERANG fields.  The WOMBAT extrapolated fluxes are converted
to temperature using a Gaussian beam that is a good approximation of the
beam + pixel window function.  Assuming that each of the WOMBAT sources
is in a separate pixel, the $rms$ contributed by these point sources to
the map is calculated.  In the $C_\ell$ power spectrum this should show
up as a constant $C_\ell = C_0$ contribution, which is found by using
the effective rms contributed by a random distribution of point sources,
$rms^2 = \sum_\ell{( 2\ell+1) C_0 W_\ell/ (4 \pi)}$, and the known beam
window function $W_\ell$.  For results quoted in the units of
Table~\ref{table:spec}, this leads to an estimated point source
contribution as a function of $\ell$ of $160 (\ell/1000)^2 \mu$K$^2$.

However, three quasars are easily identified in
the maps and removed.  The brightest two of these are 
also the two highest flux objects in the Wombat catalog in our region;
the third quasar is the eighth brightest in the catalog.  Removing only
the two brightest sources from the catalog and repeating the
above analytic estimate leads to a prediction for the contribution
of the remaining sources of only $85(l/1000)^2 \mu$K$^2$.

The power spectrum is evaluated directly from the maps before and after
removing the three quasars.  This was done by ignoring pixels within
$0.5^o$ of the quasar position.  This cutting induces very small
additional bin-bin correlations in the power spectrum, which are
negligible given the small area of the cuts.  The effect of cutting the
three quasars is less than $170 \mu$K$^2$ at all $\ell<1000$.  This, combined
with the analytic estimates above, gives us good confidence that the
residual radio point source contamination is far less than the quoted
errors at all $\ell$.

 \section{Power Spectra}

\begin{figure}[tbp]
 \plotone{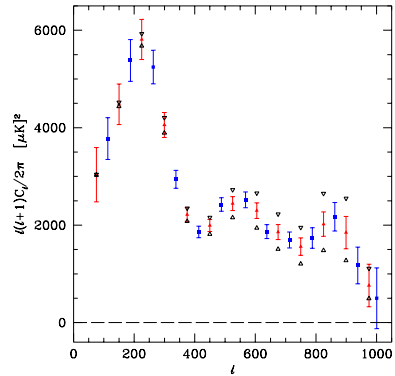} \caption{The angular power spectrum of the
 CMB, as measured at 150 GHz by BOOMERANG.  The vertical error bars show
 the statistical + sample variance errors on each point.  There is a
 common 10\% calibration uncertainty in temperature, which becomes a
 20\% uncertainty in the units of this plot.  The points are also
 subject to an uncertainty in the effective beam width of $\pm1.4'$ ($1
 \sigma$).  The effect of a $1\sigma$ error in the beam width would be to
 move the red points (all up together if the beam width has been
 underestimated, or all down together if the beam width has been
 overestimated) to the positions shown by the black triangles.  The blue
 points would move in a similar fashion.  The blue and red points show
 the results of two independent analyses using top-hat binnings offset and
 overlapping by 50\%.  This shows the basic result is not dependent on
 binning.  While each of the independent spectra (red circles or blue
 squares) are internally nearly uncorrelated, each red point is highly
 correlated with its blue neighbors, and vice versa.  These data are
 listed in Table~\ref{table:spec75}}
\label{fig:spec75}  
\end{figure}

\begin{deluxetable}{ccccc}
\tabletypesize{\scriptsize}
\tablecaption{The Angular Spectrum \label{table:spec}}
\tablewidth{0pt}
\tablehead{ 
\colhead{$\ell$ range} & \colhead{$\frac{\ell(\ell+1)}{2\pi}C_\ell (\mu
K^2)$}
& \colhead{(1dps-2dps)/2 ($\mu K^2$)} & \colhead{(left-right)/2 ($\mu K^2$)} &
\colhead{((A+A1)-(A2+B2))/2 ($\mu K^2$)}}
\startdata
(~76~ - ~125) & $3519~\pm~558$ & $\phm{-}\phn\phn2~\pm~\phn\phn8$ &
$-\phn\phn11~\pm~\phn\phn11$ & $\phm{-}\phn\phn4~\pm~\phn\phn4$ \\
(126~ - ~175) & $4688~\pm~555$ & $\phm{-}\phn40~\pm~\phn13$ &
$-\phn\phn17~\pm~\phn\phn13$ & $\phm{-}\phn\phn3~\pm~\phn\phn4$ \\
(176~ - ~225) & $5592~\pm~548$ & $\phm{-}\phn23~\pm~\phn14$ &
$-\phn\phn15~\pm~\phn\phn19$ & $-\phn\phn1~\pm~\phn\phn5$ \\
(226~ - ~275) & $5699~\pm~486$ & $\phm{-}\phn45~\pm~\phn19$ &
$-\phn\phn27~\pm~\phn\phn25$ & $\phm{-}\phn\phn0~\pm~\phn\phn7$ \\
(276~ - ~325) & $3890~\pm~316$ & $\phm{-}\phn69~\pm~\phn24$ &
$-\phn\phn44~\pm~\phn\phn33$ & $-\phn\phn8~\pm~\phn\phn9$ \\
(326~ - ~375) & $2591~\pm~207$ & $\phm{-}\phn10~\pm~\phn26$ &
$-\phn\phn32~\pm~\phn\phn44$ & $-\phn\phn9~\pm~\phn12$\\
(376~ - ~425) & $1842~\pm~152$ & $\phm{-}\phn14~\pm~\phn33$ &
$-\phn\phn98~\pm~\phn\phn55$ & $-\phn\phn3~\pm~\phn16$\\
(426~ - ~475) & $2070~\pm~161$ & $-\phn58~\pm~\phn37$ &
$-\phn\phn90~\pm~\phn\phn70$ & $-\phn\phn8~\pm~\phn20$ \\
(476~ - ~525) & $2267~\pm~174$ & $\phm{-}\phn24~\pm~\phn53$ &
$-\phn\phn30~\pm~\phn\phn95$ & $\phm{-}\phn23~\pm~\phn27$ \\
(526~ - ~575) & $2293~\pm~182$ & $-\phn\phn9~\pm~\phn68$ &
$-\phn150~\pm~\phn118$ & $\phm{-}\phn26~\pm~\phn36$ \\
(576~ - ~625) & $2058~\pm~181$ & $\phm{-}100~\pm~\phn93$ &
$-\phn161~\pm~\phn155$ & $-\phn11~\pm~\phn43$ \\
(626~ - ~675) & $1934~\pm~190$ & $\phm{-}\phn28~\pm~115$ &
$\phm{-}\phn203~\pm~\phn217$ & $-\phn23~\pm~\phn56$ \\
(676~ - ~725) & $1828~\pm~207$ & $-\phn58~\pm~145$ &
$\phm{-}\phn\phn71~\pm~\phn269$ & $-\phn32~\pm~\phn71$ \\
(726~ - ~775) & $1440~\pm~226$ & $\phm{-}196~\pm~198$ &
$-\phn421~\pm~\phn324$ & $\phm{-}\phn89~\pm~\phn99$ \\
(776~ - ~825) & $1920~\pm~288$ & $-336~\pm~235$ & $-\phn808~\pm~\phn411$ &
$\phm{-}160~\pm~131$ \\
(826~ - ~875) & $2243~\pm~361$ & $-211~\pm~317$ &
$-\phn\phn73~\pm~\phn580$ & $\phm{-}176~\pm~171$ \\
(876~ - ~925) & $1752~\pm~428$ & $-\phn94~\pm~437$ &
$-\phn613~\pm~\phn757$ & $-\phn23~\pm~217$ \\
(926~ - ~975) & $~985~\pm~506$ & $-\phn78~\pm~591$ & $-\phn607~\pm~1013$ &
$-458~\pm~278$ \\
(976~ - 1025) & $~502~\pm~627$ & $-128~\pm~800$ & $-1370~\pm~1347$ &
$-\phn82~\pm~395$ \\
\enddata


\tablecomments{The spectrum of the CMB, as used in the parameter
extraction.  The spectrum is further subject to an overall 10\%
calibration uncertainty, and a 1.4' effective beam uncertainty.  The
spectrum of the all-channel consistency tests are also given.  Adjacent
bins are weakly correlated}

\end{deluxetable}


\begin{deluxetable}{cc}
\tabletypesize{\scriptsize}
\tablecaption{The Angular Spectrum \label{table:spec75}}
\tablewidth{0pt}
\tablehead {
\colhead{$\ell$ range} & \colhead{$\frac{\ell(\ell+1)}{2\pi}C_\ell (\mu
K^2)$}}
\startdata
(~50~ - ~113) & $3035~\pm~557$\\
(~75~ - ~150) & $3776~\pm~428$\\
(112~ - ~187) & $4481~\pm~416$\\
(150~ - ~225) & $5380~\pm~429$\\
(187~ - ~262) & $5810~\pm~413$\\
(225~ - ~300) & $5245~\pm~345$\\
(262~ - ~337) & $4056~\pm~257$\\
(300~ - ~375) & $2942~\pm~184$\\
(337~ - ~412) & $2218~\pm~140$\\
(375~ - ~450) & $1861~\pm~119$\\
(412~ - ~487) & $1992~\pm~123$\\
(450~ - ~525) & $2424~\pm~138$\\
(487~ - ~562) & $2443~\pm~142$\\
(530~ - ~605) & $2520~\pm~162$\\
(567~ - ~642) & $2298~\pm~160$\\
(600~ - ~675) & $1868~\pm~144$\\
(637~ - ~712) & $1858~\pm~154$\\
(675~ - ~750) & $1696~\pm~163$\\
(712~ - ~787) & $1560~\pm~179$\\
(750~ - ~825) & $1736~\pm~211$\\
(787~ - ~862) & $2021~\pm~250$\\
(825~ - ~900) & $2172~\pm~292$\\
(862~ - ~937) & $1847~\pm~333$\\
(900~ - ~975) & $1174~\pm~377$\\
(937~ - 1012) & $~762~\pm~437$\\
(963~ - 1038) & $~499~\pm~623$\\
\enddata


\tablecomments{The spectrum of the CMB, as shown in
Figure~\ref{fig:spec75}.  These are the the results of two independent
analyses using $\Delta l = 75$ top-hat binnings offset by $\Delta l =
75/2$ and thus overlapping by 50\%.  While each of the independent
spectra are internally nearly uncorrelated, each red point is highly
correlated with its neighbors from the other binning.  This binning is
not used in the parameter extraction.  Rather, the non-overlapping (and
thus only weakly correlated) $\Delta l = 50$ top-hat binning listed in
Table~\ref{table:spec} is.  The spectrum is further subject to an
overall 10\% calibration uncertainty, and a 1.4' effective beam
uncertainty.}

\end{deluxetable}

The results are summarized in Table~\ref{table:spec75} and
Table~\ref{table:spec}, and in Figure~\ref{fig:spec75} and
Figure~\ref{fig:models}.  

The sensitivity of the results to different
binnings is explored.  Figure~\ref{fig:spec75} and
Table~\ref{table:spec75} summarize the results from two independent
analysis using top-hat binning of width $\Delta l = 75$, and offset and
overlapping by 50\%.  Because of this overlap, adgacent points are
strongly correlated.  This binning is not used in the parameter
extraction.

Figure~\ref{fig:models} and Table~\ref{table:spec} summarized the
results from an analysis with non-overlapping top-hat bins of width
$\Delta l = 50$.  The sources of uncertainty that are included in the
errors quoted in Table~\ref{table:spec} include sample variance and
statistical noise. The former dominates at $\ell < \approx 600$ and the
latter at higher $\ell$.  These uncertainties are only weakly
correlated.

Uncertainty in the effective beam-size introduces an additional
uncertainty in the power spectrum that is highly correlated across the
spectrum.  The uncertainty in the effective beam size has contributions
from uncertainty in the physical beam and from uncertainty in the rms
amplitude of the pointing jitter.  These combine to produce an
uncertainty in the effective beam of $\pm13\%$.  This uncertainty is not
included in the errors quoted in Table~\ref{table:spec}, as its effect is
to produce an overall tilt to the spectrum.  The amplitude of the tilt
corresponding to the 1 sigma uncertainty that is assigned to the effective
beam width is illustrated in Figure~\ref{fig:spec75}.  This uncertainty is
included in the parameter estimation outlined in the next section.

Uncertainty due to instrumental and atmospheric artifacts in the maps
are small, based on the internal consistency tests, and are neglected.
Similarly, contamination of the maps by both diffuse and compact
astrophysical foregrounds are also negligible with respect to the other
uncertainties and are neglected.

\section{Cosmological Parameters}

\begin{figure}[tbp]
  \plotone{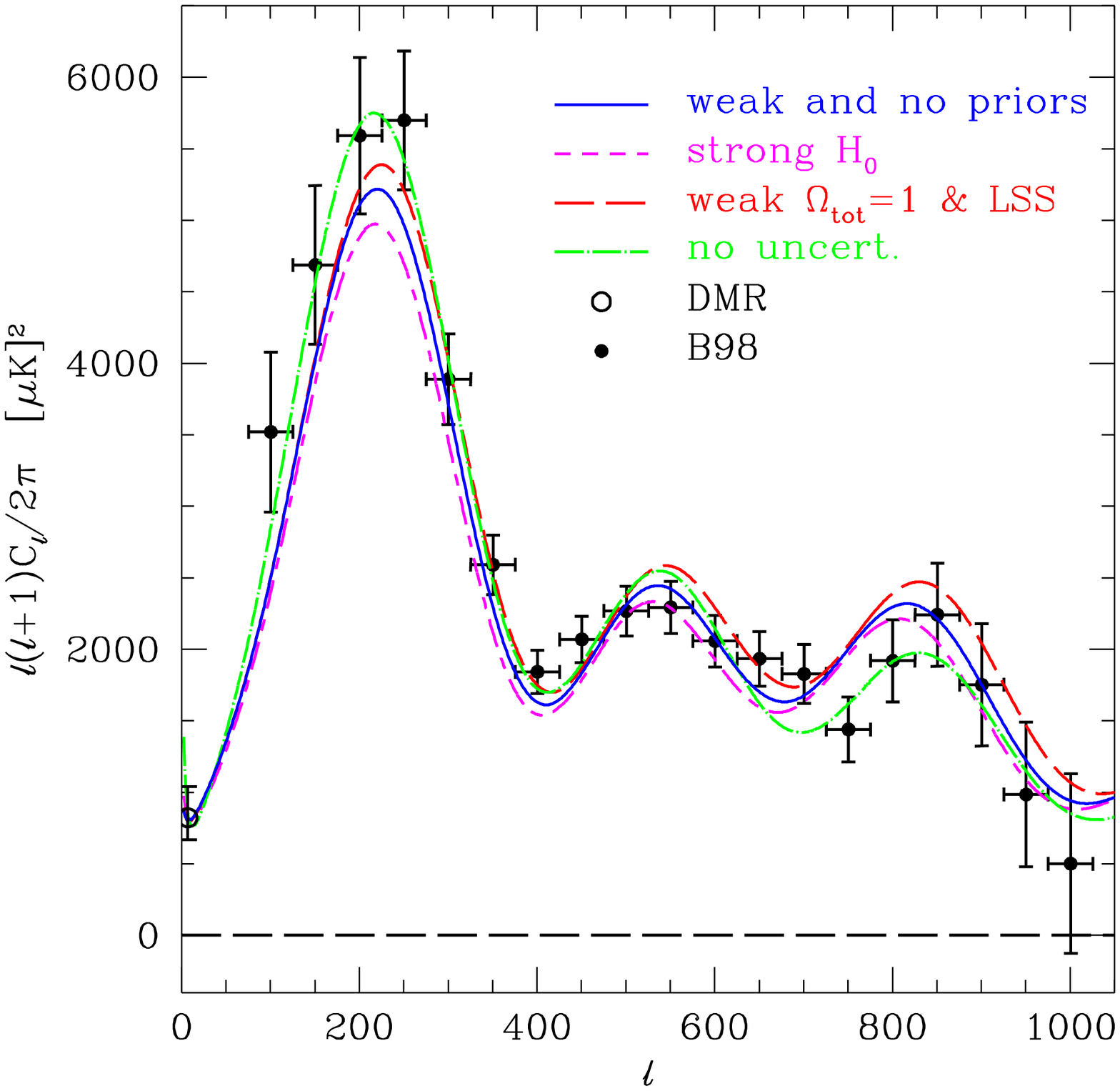}\\ \plotone{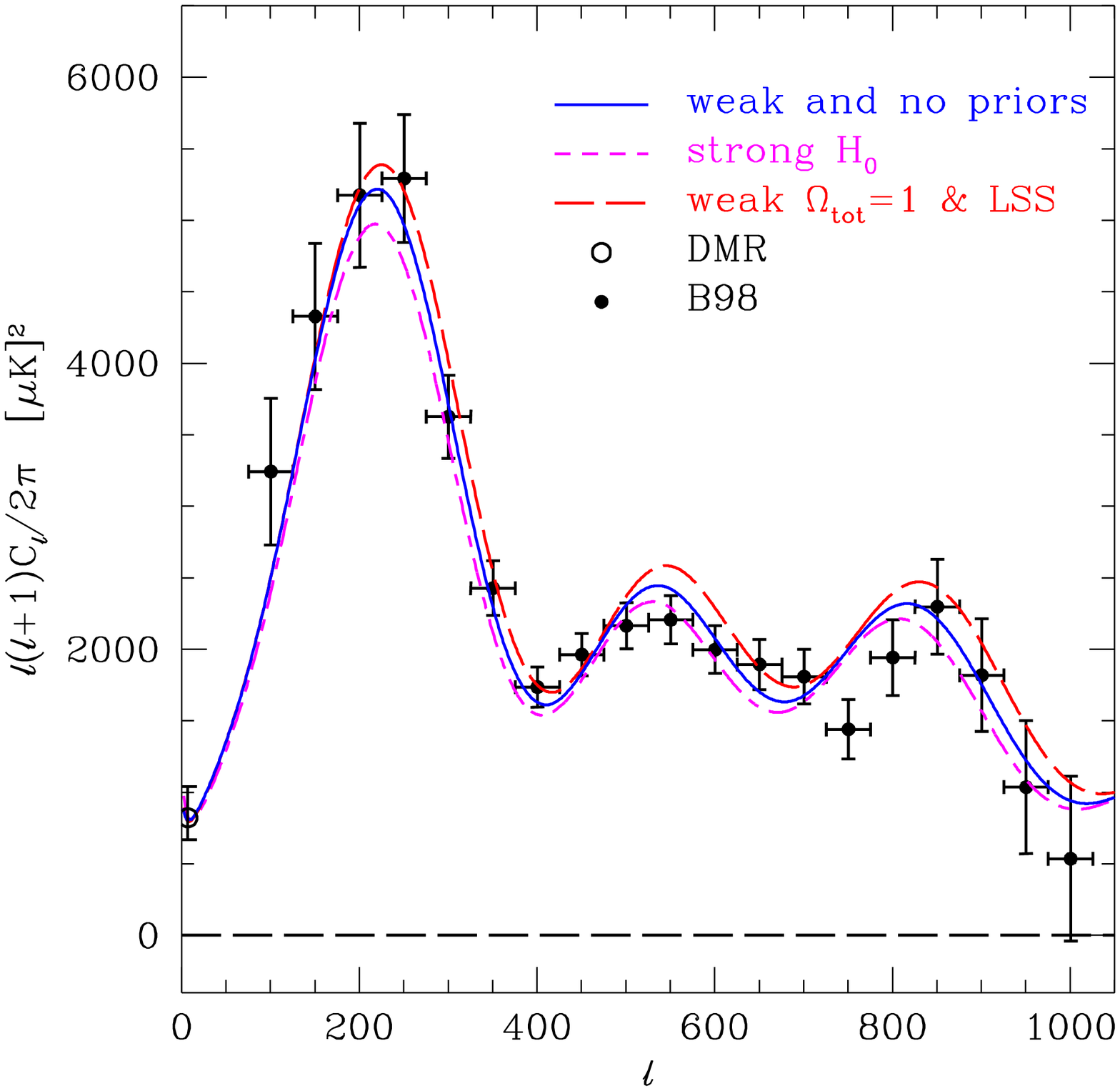} \caption{\scriptsize Selected best fit models
  normalized to the best compromise amplitude between COBE-DMR and
  BOOMERANG are shown overlayed on the BOOMERANG spectrum. The upper
  panel shows the points plotted as listed in
  table~\ref{table:spec}. The best-fit models for the ``weak'' and ``no
  priors'' cases coincide (blue, solid curve) with $\Omega_{tot} = 1.05,
  H_0 = 50, \Omega_{\Lambda} = 0.5, \omega_b = 0.020, \omega_c = 0.120,
  \tau_c= 0, n_s=0.925, t_0 = 15.8 Gyrs $. Strong Hubble prior gives the
  best fit model with parameters $\Omega_{tot} = 1., H_0 = 68,
  \Omega_{\Lambda} = 0.7, \omega_b = 0.020, \omega_c = 0.120, \tau_c= 0,
  n_s=0.925, t_0 = 13.8 Gyrs $.  The full analysis takes into account
  the calibration and beam uncertainties which best fit models take
  advantage of. This explains the apparent mismatch between some of the
  models in the upper panel and the plotted central values of Boomerang
  band powers.  The green (dash-dot) curve is the best fit model
  ($\Omega_{tot} = 1.15, H_0 = 42, \Omega_{\Lambda} = 0.7, \omega_b =
  0.020, \omega_c = 0.060, \tau_c= 0.2, n_s=0.925, t_0 = 20 Gyrs$) when
  both beam and calibration uncertainties are switched off.  The model
  fits closely the central values as expected.  To demonstrate the
  effect of beam and calibration uncertanties, in the lower panel the
  data points have been replotted with a $4\%$ decrease in calibration
  ($0.4 \sigma$) and a 0.5 arcminute change in beam size ($0.4 \sigma$).
  The plot makes it clear that the best-fit conventional CDM models are
  indeed good fits to the data, once these uncertainties are correctly
  accounted for.}
\label{fig:models}  
\end{figure}

The angular power spectrum shown in Figure~\ref{fig:models} can be used
in conjunction with other information to determine cosmological
parameters.  In this paper, the parameter extraction methods and tools
described in detail in \citet{lange01} are used.

Specifically, the relative agreement between these data and theoretical
predictions over a broad 7-dimensional cosmological parameter space is
explored.  This parameter space is appropriate for models with adiabatic
initial conditions (e.g. inflationary models).  The resolution of the
second peak virtually rules out alternative models such as defect based
scenarios which predict a single broader peak with no secondaries
\citep{Turo98,Cont99}.

\begin{figure}[tbhp]
  \plotone{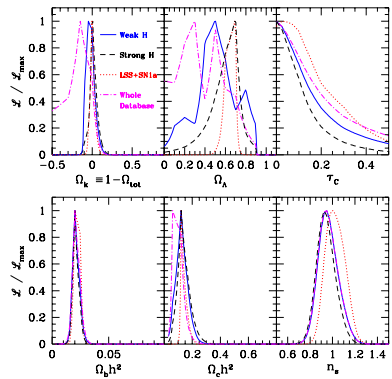} \caption{ Likelihood functions for a subset of the
  priors used in Table~\ref{table:parameters}. $\Omega_bh^2$, and $n_s$
  are well constrained, even under the ``whole database'' case, and are
  insensitive to additional priors.  $\Omega_{\rm tot}$ is poorly
  constrained over the whole database, but when the weak priors are
  applied, it becomes stably consistent with the flat case.  With the
  weak priors, $\Omega_\Lambda$ and $\Omega_c h^2$ are poorly
  constrained, but become significant detections with the addition of
  the other priors considered.  For all cases, $\tau_c$ is poorly
  constrained, but does prefer low values.} \label{fig:like}
\end{figure}

Parameters explored include those describing energy densities, including
the total energy density $\Omega_{\rm {tot}}$, the vacuum energy density
$\Omega_\Lambda$, and the physical densities of baryons and cold dark
matter, $\Omega_b h^2$ and $\Omega_c h^2$ respectively.  The power
spectrum of initial adiabatic density fluctuations is described by a
normalization $\ln\mathcal{C}_{10}$ and a power law exponent $n_s$.  The
effects of recent reionization of the universe on the observed angular
power spectrum, parameterized by the optical depth to decoupling,
$\tau_C$, are also explored. For this parameter, the liklihood does not
fall sufficiently by the edge of the explored parameter range
($\tau_c<0.5$) to produce significant limits, though a preference for
low values of $tau_c$ are evident in Figure~\ref{fig:like}.


\begin{deluxetable}{llllllllll}
\tabletypesize{\scriptsize}
\tablecaption{Results of Parameter Extraction\label{table:parameters}}
\tablewidth{0pt}
\tablehead{\colhead{Priors}
& \colhead{$\Omega_{tot}$}
& \colhead{$n_s$}
& \colhead{$\Omega_bh^2$}
& \colhead{$\Omega_{cdm}h^2$}
& \colhead{$\Omega_{\Lambda}$}
& \colhead{$\Omega_m$}
& \colhead{$\Omega_b$}
& \colhead{$h$}
& \colhead{Age}
}
\startdata
Weak only
& $1.02^{0.06}_{0.06}$ 
& $0.96^{0.10}_{0.09}$ 
& $0.022^{0.004}_{0.003}$ 
& $0.13^{0.05}_{0.05}$ 
& $(0.51^{0.23}_{0.20})$ 
& $(0.51^{0.20}_{0.20})$ 
& $0.07^{0.03}_{0.03}$ 
& $(0.56^{0.10}_{0.10})$ 
& $15.2^{1.9}_{1.9}$ 
\\
\tableline
LSS
& $1.02^{0.04}_{0.05}$ 
& $0.97^{0.10}_{0.08}$ 
& $0.022^{0.004}_{0.003}$ 
& $0.13^{0.03}_{0.02}$ 
& $0.55^{0.09}_{0.09}$ 
& $0.49^{0.12}_{0.12}$ 
& $0.07^{0.02}_{0.02}$ 
& $0.56^{0.09}_{0.09}$ 
& $15.0^{1.3}_{1.3}$ 
\\
SN1a
& $1.02^{0.07}_{0.05}$ 
& $0.99^{0.11}_{0.10}$ 
& $0.023^{0.004}_{0.004}$ 
& $0.10^{0.04}_{0.04}$ 
& $0.73^{0.07}_{0.10}$ 
& $0.31^{0.06}_{0.06}$ 
& $0.06^{0.03}_{0.03}$ 
& $0.61^{0.09}_{0.09}$ 
& $15.9^{2.5}_{2.5}$ 
\\
LSS \& SN1a
& $0.99^{0.03}_{0.04}$ 
& $1.03^{0.10}_{0.09}$ 
& $0.023^{0.003}_{0.003}$ 
& $0.14^{0.03}_{0.02}$ 
& $0.65^{0.05}_{0.06}$ 
& $0.34^{0.07}_{0.07}$ 
& $0.05^{0.02}_{0.02}$ 
& $0.67^{0.09}_{0.09}$ 
& $13.7^{1.2}_{1.2}$ 
\\
$h = 0.71 \pm 0.08$
& $0.98^{0.04}_{0.05}$ 
& $0.97^{0.10}_{0.09}$ 
& $0.022^{0.004}_{0.003}$ 
& $0.14^{0.05}_{0.04}$ 
& $0.62^{0.10}_{0.18}$ 
& $0.40^{0.13}_{0.13}$ 
& $0.05^{0.02}_{0.02}$ 
& $(0.65^{0.08}_{0.08})$ 
& $13.7^{1.6}_{1.6}$ 
\\
\tableline
Flat
& (1.00) 
& $0.95^{0.09}_{0.08}$ 
& $0.021^{0.003}_{0.003}$ 
& $0.13^{0.04}_{0.04}$ 
& $(0.57^{0.12}_{0.37})$ 
& $(0.48^{0.24}_{0.24})$ 
& $0.06^{0.02}_{0.02}$ 
& $(0.61^{0.13}_{0.13})$ 
& $14.3^{0.6}_{0.6}$ 
\\
Flat \& LSS
& (1.00) 
& $0.98^{0.10}_{0.07}$ 
& $0.021^{0.003}_{0.003}$ 
& $0.13^{0.01}_{0.01}$ 
& $0.62^{0.07}_{0.07}$ 
& $0.38^{0.07}_{0.07}$ 
& $0.05^{0.01}_{0.01}$ 
& $0.62^{0.06}_{0.06}$ 
& $14.5^{0.7}_{0.7}$ 
\\
Flat \& SN1a
& (1.00) 
& $0.98^{0.11}_{0.09}$ 
& $0.022^{0.003}_{0.003}$ 
& $0.12^{0.01}_{0.02}$ 
& $0.68^{0.04}_{0.06}$ 
& $0.33^{0.05}_{0.05}$ 
& $0.05^{0.01}_{0.01}$ 
& $0.66^{0.05}_{0.05}$ 
& $14.0^{0.6}_{0.6}$ 
\\
Flat, LSS \& SN1a
& (1.00) 
& $1.03^{0.10}_{0.09}$ 
& $0.023^{0.003}_{0.003}$ 
& $0.13^{0.01}_{0.01}$ 
& $0.66^{0.04}_{0.06}$ 
& $0.33^{0.05}_{0.05}$ 
& $0.05^{0.01}_{0.01}$ 
& $0.66^{0.05}_{0.05}$ 
& $14.0^{0.6}_{0.6}$ 
\\
\enddata
\tablecomments{Results of parameter extraction using successively more
  restrictive priors, following \citet{lange01}.  The confidence
  intervals are 1$\sigma$.  The quoted values are reported after
  marginalizing over all other parameters.  For the primary database
  parameters, 16\% and 84\% integrals are reported as $\pm1\sigma$
  errors. For $\Omega_m$, $\Omega_b$, $h$, and Age, which are functions
  of the other parameters, the mean and standard deviation over the
  distribution are reported. All entries are subject to a weak prior in
  which only models with $0.45<h<0.90$ and age~$> 10$~Gyr are
  considered.  The LSS \citep{bj99} and SN1a supernovae \citep{riess,
  perlm} priors are as described in \citet{lange01}.  The strong $h$
  prior is a Gaussian with the stated 1$\sigma$ error. Parentheses are
  used to indicate parameters that did not shift more than 1-$\sigma$ or
  have their errors reduced by a factor of two upon the inclusion of the
  CMB data, compared with an analysis using the priors only.  Thus, in
  these cases the parameter range reflects the choice of prior, rather
  than a constraint by the CMB.  The age column is in units of Gyr.}
\end{deluxetable}

Given the data, likelihoods as a function of theoretically predicted
power spectra are calculated throughout this parameter space.  For every
comsological model, beam and calibration uncertainties add two
additional parameters. We approximate the possible correction to the
beam of effective width $\omega$ by a gaussian form $W_\ell
(\omega+\delta \omega)/W_\ell(\omega)= e^{-\ell(\ell+1) \omega \delta
\omega} $, with $\delta \omega$ assumed to be gaussian distributed with
the standard deviation corresponding to $1.4'$. Effectively, every theoretical
spectrum is multipled by $e^{2 \ell(\ell+1) \omega \delta \omega} $.
Calibration uncertainty of 10\% adds to the variance of
$\ln\mathcal{C}_{10}$.  We do not calculate likelihoods on a grid in
beam-width -- calibration space, rather for every model we search for a
maximum of likelihood in $\delta \omega$ and $\ln\mathcal{C}_{10}$,
calculate the curvature of the likelihood near this maximum, and
marginalize over the beam and the amplitude by integrating likelihood in
the gaussian approximation.  The best fit value of $\ln\mathcal{C}_{10}$
and it's variance is used when combining Boomerang predictions with
other data.

Parameters are constrained individually by marginalizing over all
others, including two describing the calibration and beam window
function uncertainties.  The 16\% and 84\% integrals are reported as
$\pm1\sigma$ errors. Other quantities such as the Hubble constant and
the age of the Universe are derived from those used to define the
parameterization, using the mean and variance over the posterior
distribution.  The details of the discrete numerical database used for
this process, including the limits and values of each parameter tested,
and exact prescription used for calculating likelihoods and extracting
confidence limits, is fully described in \citet{lange01}.

Before marginalization, the calculated likelihood for each model is
multiplied by the likelihood derived from a series of ``prior
probabilities'', or priors, which represent knowledge from other
cosmological measurements.  All results considered here have a
``weak $h$ + age'' top-hat prior applied (hereafter simply the ``weak
prior'') which eliminates models where the universe is younger than
10~Gyr, and limits the Hubble constant, $H_0 = 100 \, h \mbox{\, km
s}^{-1}\mbox{Mpc}^{-1}$, to $0.45 \le h \le 0.9$.

Applying stronger priors in conjunction with this weak prior exercises
the ability of the CMB data to combine with other measurements (or
theoretical prejudice) to significantly narrow the parameter confidence
 intervals. Considered here is the impact of applying a stronger
constraint on $h$, constraints derived from measurements of large scale
structure (LSS)\citep{bj99}, results from recent measurements of type Ia
supernovae~\citep{riess, perlm}, and the theoretical bias that
$\Omega_{\rm tot} = 1$.

The parameter estimates, given these various combinations of priors, 
are shown in Table~\ref{table:parameters}, with marginalized likelihood
curves for several important parameters given in
Figure~\ref{fig:like}.

\citet{lange01} reports a family of models within the database that
provide good fits to the angular power spectrum up through $\ell \sim
600$, but represent very young, high baryon density, very closed models
outside the normal realm of consideration in modern cosmology.  In
\citet{lange01}, the weak prior was used to keep these models from
affecting the parameter estimates.  Using the power spectrum shown in
Figure~\ref{fig:models}, but limiting consideration to the points with
$\ell \le 600$, similar behavior is exhibited; the very young, high
baryon density, very closed models dominate the fits, but can be
eliminated by the weak prior.  As in \citet{Jaf01}, adding the higher
$\ell$ points, which exclude models with either a very high or very
small third peak, eliminates the high baryon density models. It does
not eliminate the closed models in the absence of the weak priors.
Thus, just as in \citet{lange01}, cases with the weak priors applied are
the focus of the extraction.

As is apparent in Figure~\ref{fig:like} and
Table~\ref{table:parameters}, $\Omega_b h^2$ and $n_s$ are well
localized for all choices of priors.  The range for $\Omega_b h^2$ is
very consistent with determinations based on Big Bang Nucleosynthesis
and measurements of light element abundances \citep{Tyt00}, while $n_s$
is localized near unity, consistent with inflation-based models.
Similarly, once the weak priors are applied, $\Omega_{\rm tot}$ is well
constrained and consistent with a flat universe.
$\Omega_{\Lambda}$ and $\Omega_c h^2$ are poorly constrained for the
weak prior case.  $\tau_C$ favors low values, but not at a
statistically significant level.

These limits agree well with those found in \citet{lange01}, with one
exception.  There, the $1\sigma$ range for $\Omega_b h^2$ (with the weak
prior) is $0.036\pm0.005$.  Considering only the points in
Figure~\ref{fig:models} with $\ell \le 600$ results in a $1\sigma$ limit of
$0.027\pm0.005$.  This shift is presumably due to the improvement of the
pointing solution and smaller error bars compared with B00.
The table reflects the estimate made using all points up to
$\ell = 1000$;  the addition of the information contained in the
high-$\ell$ points has moved the confidence intervals still further.

Having found tight limits on $\Omega_b h^2$, $\Omega_{\rm tot}$, and $n_s$
with the weak priors, the impact of adding other priors can be
discussed, to see whether these results are stable (i.e., consistent
with the prior) and whether any new
parameters can be localized.  The table indicates that, in fact, these
three parameters are very stable to the addition of the LSS, SN1a,
strong $h$ and the theoretically motivated $\Omega_{\rm tot} = 1$ priors, in
various combinations.  This insensitivity to choice of priors is a
powerful indication that the cosmology is consistent and that these
parameters have been robustly determined.

The table also shows that the CMB data can be combined with these priors
independently to make statistically significant determinations of
$\Omega_{\Lambda}$ and $\Omega_c h^2$.  While the confidence intervals
shift somewhat depending on the chosen prior, the rough agreement among
these three results, giving $\Omega_{\Lambda} \sim 0.65$ and $\Omega_c
h^2 \sim 0.12$, is very compelling.  These LSS and SN1a results are
similar to those found in \citet{lange01}.  For the first time, however,
the combination of CMB data with just a strong limit on $h$ is powerful
enough to yield such a detection.

The table also shows a consistent story for the age of the universe and
for the Hubble Constant.  For the prior combinations where the limits on
$h$ are dominated by the CMB data rather than the priors, the extraction finds
$h \sim 0.65$, with ages between 13 and 15 Gyr.  These quantities are
most strongly constrained by the CMB data along with the combination of
the SN1a prior and flatness.

\section{Conclusions}

A measurement of the angular power spectrum of the CMB, characterized by
a series of harmonic peaks, has been presented, confirming the existence
of this unique signature of acoustic oscillations in the early universe.
This is an important confirmation of standard adiabatic
models of structure formation, and thus of the process of constraining
cosmological parameters based on increasingly precise measurements of
the position and amplitude of these peaks by current and future CMB
experiments.

The precision and extent of the angular power spectrum that is reported
here already allow an accurate determination of several cosmological
parameters with the assumption of only weak astrophysical priors.  
Assuming $0.45 < h < 0.90$, the CMB data tightly
constrain the values of $\Omega_{\rm {tot}}$ and $n_s$ to lie close to
unity and tightly constrain $\Omega_b$ to a value consistent with BBN.

Adding constraints from observations of Large Scale Structure and of
type 1a supernovae, yields a value for the Hubble constant of $h=0.67
\pm 0.09$, that is in good agreement with the HST key project final
value of $0.72 \pm 0.08$ \citep{free}.  These data also
provide compelling evidence for the existence of both dark matter and
dark energy.  Including LSS, SN1a or a prior on the Hubble constant of
$h = 0.71 \pm 0.08$ each yields $\Omega_\Lambda \approx 2/3$ and
$\Omega_m \approx 1/3$.

\acknowledgments

The BOOMERanG project has been supported by NASA and by NSF OPP in the
U.S., by PNRA, Universit\'a ``La Sapienza'', and ASI in Italy, by PPARC
in the UK, and by the CIAR and NSERC in Canada.  We received excellent
logistical support from Kathy Deniston, and superb field and flight
support from NSBF and the USAP personnel in McMurdo.


\begin{thebibliography}{}

\bibitem[Acbar, 2001]{acbar}Preliminary analysis of Acbar data, J. Ruhl, 2001,
	private communication.

\bibitem[Bond and Efstathiou(1987)]{Bon87} Bond, J. R. and Efstathiou, G.,
\mnras, 226, 655(1987). 

\bibitem[Bennet et al.(1996)]{bennet} Bennett, C.L. et al. 1996, 
	\apjl, 464, 1

\bibitem[Bond, Jaffe, and Knox(2000)]{bjk00} J.R.~Bond, A.H.~Jaffe \&
L.~Knox, \apj, 533, 19-37, 2000.  astro-ph/9808264

\bibitem[Bond, Jaffe, and Knox(1998)]{bjk98} J.R.~Bond, A.H.~Jaffe \&
L.~ Knox, \prd, 57, 2117, 1998. astro-ph/9708203

\bibitem[Bond \& Jaffe(1999)]{bj99} J.R. Bond and A.H. Jaffe,
Phil. Trans. R.  Soc. London, 357, 57(1999), astro-ph/9809043.  

\bibitem[Bond et al.(2001)]{b01}Bond, J.R. et al., 2001, in preparation.

\bibitem[Borrill(1999)]{Borr1999} 
Borrill J. 1999, Proc. of the 3K Cosmology EC-TMR conference,  
eds. L. Maiani, F. Melchiorri, N. Vittorio, AIP CP 476, 277

\bibitem[Church et al.(1997)]{church} Church, S.E., Ganga, K.M., Ade,
P.A.R., Holzapfel, W.L., Mauskopf, P.D., Wilbanks, T.M. and Lange, A.E.
1997, ApJ, 484, 523 

\bibitem[Contaldi et al.(1999)]{Cont99} Contaldi, C.R., Hindmarsh,
M.B. and Magueijo, J., 1999, Phys. Rev. Lett. 82 679-682

\bibitem[Crill et al.(2001)]{Cril2001} Crill, B. et al., 2001, in
preparation

\bibitem[de Bernardis et al.(2000)]{debe2000} de Bernardis, P., et
al. 2000, Nature, 404, 955-959

\bibitem[de Bernardis et al. (2002)]{deb01} de Bernardis, P., et al,
2002, \apj, 564.

\bibitem[Efstathiou and Bond(1999)]{Efs99} Efstathiou G., and Bond, J. R., 
\mnras, 304, 75(1999).

\bibitem[Finkbeiner et al.(1999)]{Fink99} Finkbeiner D.P. et al. 1999,
\apj, 524, 867.

\bibitem[Freedman et al, 2000]{free} Freedman W. L. et al, 2000, ApJ in
press, preprint astro-ph/0012376.

\bibitem[G\'orski et al.(1998)]{Gors1998} G\'orski, K.M., Hivon, E. and
Wandelt, B.D., in "Analysis Issues for Large CMB Data Sets", 1998,
eds. A.J. Banday, R.K. Sheth and L. Da Costa, ESO(astro-ph /9812350),
see also http://www.tac.dk/~healpix/

\bibitem[Jaffe et al.(2001)]{Jaf01} Jaffe, A., et al., 2001, \prl, 86, 3475-3479

\bibitem[Hanany et al.(2000)]{Hana2000} Hanany, S. et al., 2000, \apj,
545, L5-L9 

\bibitem[Hivon et al.(2001)]{Hivo2001} Hivon, E., Gorski, K.M.,
Netterfield, C.B., Crill, B.P., Prunet, S. \& Hansen F., 2001,
astro-ph/0105302, accepted in ApJ 

\bibitem[Hu et al.(1997)]{Hu1997} Hu W., Sugiyama N. \& Silk J., 1997,
Nature, 386, 37

\bibitem[James(1981)]{james} James,G.L.\lq\lq Analysis and design of TE11
to HE11 corrugated cylindrical waveguide mode converters\rq\rq, {\em
IEEE Trans. Microwave Theory and Techniques}, Vol. MTT-29, pp 1059-1066,
1981.

\bibitem[Lange et al.(2001)]{lange01} Lange, A.E., et al, 2001, 
\prd, 63, 042001

\bibitem[Lineweaver et al.(1996)]{Line1996} Lineweaver, C.H., et al. 1996,
\apj, 470:38-42.

\bibitem[Masi et al.(2001)]{Mas01} Masi S., et al. 2001, \apjl in
press, astro-ph/0101539

\bibitem[Mauskopf et al.(2000)]{Maus2000} Mauskopf, P. et al. 2000, \apj,
536, L59

\bibitem[Miller et al.(1999)]{Mil99} Miller, A. et al. 1999, \apj, 524,
L1

\bibitem[Murphy et al.(2001)]{murphy} Murphy, J.A.,  Ruth Colgan,
Creidhe O\rq Sullivan, Bruno Maffei, and Peter Ade.  "Radiation Patterns
of Multi-Moded Corrugated Horns for Far-IR space
Applications". Preprint, TeraHz Technology Conference, 2001.

\bibitem[Netterfield et. al.(1997)]{Net97} Netterfield, C. B., et. al.
1997, \apj, 474, 47

\bibitem[Padin et al.(2001)]{Pad01} Padin, S., et al., 2001, \apj, 549,
L1-L5

\bibitem[Peebles \& Yu(1970)]{Peb70} Peebles, P.J.E, and Yu J.T., 1970,
\apj 162, 815


\bibitem[Piacentini et al.(2002)]{Piac2001}
Piacentini, F. et al., 2002, accepted by \apjs

\bibitem[Prunet et al.(2000)]{Prun2000} Prunet, S. et al., 2000, in "Energy
densities in the Universe", Bartlett J., Dumarchez J. eds., Editions
Frontieres, Paris - astro-ph/0006052 

\bibitem[Perlmutter et al.(1999)]{perlm} S. Perlmutter et al.,
\apj, 517, 565(1999)

\bibitem[Riess et al.(1998)]{riess} Riess et al., 1998, \aj, 116, 1009

\bibitem[Rao et al.(1982)]{rao} Rao,~S.M.,  D.~R.~Wilton, and
A.~W.~Glisson,\lq\lq Electromagnetic scattering by surfaces of arbitrary
shape\rq\rq, {\em IEEE Trans. Antennas and Propagation}, Vol. AP-30,
no. 3, pp 409-418, May 1982.

\bibitem[Schlegel et al.(1999)]{Schl99} Schlegel D.J. et al. 1999,
\apj 500, 525. 

\bibitem[Stumpff(1980)]{Stump1980} Stumpff, 1980, A\&A Suppl, 41, 1.

\bibitem[Sunyaev \& Zeldovich(1970)]{SZ70}Sunyaev, R.A. \& Zeldovich ,
Ya.B., 1970,  Astrophysics and Space Science 7, 3-19

\bibitem[Tegmark et al.(2000)]{Tegm00} Tegmark M. et al., 2000, \apj 530,
133.

\bibitem[Turok et al.(1998)]{Turo98} Turok, N., Pen, U-L. and Seljak,
U. 1998, Phys. Rev. D58 023506

\bibitem[Tytler et al.(2000)]{Tyt00} Tytler, D. et al. 2000, \physscr,
submitted, astro-ph/0001318.

\bibitem[Wright(1998)]{Wrig98} Wright, E.L., 1998, \apj 496, 1.

\bibitem[WOMBAT(1998)]{wombat} WOMBAT collaboration, 1998, see
http://astron.berkeley.edu/wombat/foregrounds/radio.html.

\end{thebibliography}
\end{document}